\documentclass[a4paper,12pt,abstract=true]{scrartcl}
\pdfoutput=1
\usepackage{authblk}
\usepackage{amsmath,amssymb}
\usepackage{bm}
\usepackage{hyperref}
\usepackage{graphicx,color}
\RequirePackage[l2tabu, orthodox]{nag}
\usepackage[all,warning]{onlyamsmath}

\begin{document}

\titlehead{\hfill KUNS-2646, OU-HET 914}

\title{Towards Background-free RENP\\ using a Photonic Crystal Waveguide}

\author[1]{Minoru~Tanaka\thanks{Email: \texttt{tanaka@phys.sci.osaka-u.ac.jp}}}
\affil[1]{Department of Physics, Graduate School of Science, 
          Osaka University, Toyonaka, Osaka 560-0043, Japan} 

\author[2]{Koji~Tsumura}
\affil[2]{Department of Physics, Kyoto University, Kyoto 606-8502, Japan}

\author[3]{Noboru~Sasao}
\affil[3]{Research Institute for Interdisciplinary Science, 
          Okayama University, Tsushima-naka 3-1-1, Kita-ku,
          Okayama 700-8530, Japan}

\author[3]{Motohiko~Yoshimura}

\date{\normalsize February 28, 2017}

\maketitle

\begin{abstract}
We study how to suppress multiphoton emission background 
in QED against radiative emission of neutrino pair (RENP) from atoms. 
We purse the possibility of background suppression using the photonic 
band structure of periodic dielectric media, called photonic crystals. 
The modification of photon emission rate in photonic crystal waveguides
is quantitatively examined to clarify the condition of background-free
RENP.
\end{abstract}

\section{\label{Sec:Intro} Introduction}
Radiative emission of neutrino pair (RENP) is a novel process of 
atomic or molecular deexcitation in which a pair of neutrinos and 
a photon are emitted. It is shown that the spectrum of the photon
conveys information on unknown properties of neutrinos such as
absolute mass and Majorana/Dirac
distinction~\cite{DinhPetcovSasaoTanakaYoshimura2012a,YoshimuraSasao2013a}.
The rate of RENP is strongly suppressed as far as an incoherent ensemble 
of atoms is considered. In the proposed experiment~\cite{Fukumi2012a},
a rate enhancement mechanism with a macroscopic target of coherent
atoms is employed. This macrocoherent enhancement mechanism, applicable
to processes of plural particle emission, is experimentally shown to 
work as expected in a QED two-photon emission 
process~\cite{Miyamoto2014a,Miyamoto2015a}, 
paired superradiance (PSR)~\cite{YoshimuraSasaoTanaka2012a}.

In order to observe the RENP process and reveal the nature of neutrinos
in an experiment, we have to understand background processes.
It is shown that the three-photon emission process 
$|e\rangle\to|g\rangle+\gamma+\gamma+\gamma$ is also amplified
when the RENP $|e\rangle\to|g\rangle+\gamma+\nu\bar\nu$ is 
macrocoherently enhanced~\cite{YoshimuraSasaoTanaka2015a}. 
The rate of the macrocoherent QED process of three-photon emission
(McQ3) is found to be $O(10^{20})$ Hz for the transition from xenon 
$|e\rangle=6\text{s}\,^3\text{P}_1$ state of 8.437 eV excitation energy
to the ground state $|g\rangle=5\text{p}\,^1\text{S}_0$, 
while the RENP rate of the same transition is $O(10^{-3})$ Hz. 
Therefore, though reducible, the McQ3 process is a serious background
of the RENP process and the suppression of McQ3 (and similar $n$-photon 
emission processes, McQn) is mandatory.

An experimental scheme free of the McQn background is proposed
to overcome this difficulty~\cite{YoshimuraSasaoTanaka2015a}. 
The essential idea is to 
suppress the emission of background photons
in McQn process using a waveguide, like cavity 
QED~\cite{Purcell1946a,Kleppner1981a,Goy1983a,Hulet1985a}. 
There exist cutoff frequencies $\omega_c$ of the field eigenmodes in
a waveguide (of perfect conductor), which may be interpreted as 
the photon acquires a mass of $\hbar\omega_c$ in the waveguide. 
For instance, the smallest cutoff is 
$\omega_c=\pi/b\simeq 0.6\ (1\ \text{mm}/b)\ \text{meV}$ 
for a square waveguide of size $b$. 
It is proven that 
McQn is kinematically prohibited while RENP is allowed in an appropriate
region of trigger frequency if the smallest cutoff is larger than the 
neutrino mass and thus the background-free RENP is possible in principle.

The optical frequencies, however, are most relevant to
RENP signal triggered by lasers. 
Since even superconductors are far from the perfect conductor
in the optical region, a realization of the above idea is not 
straightforward. As a more realistic proposal, the use of photonic
crystal is mentioned in Ref.~\cite{YoshimuraSasaoTanaka2015a}. 
A photonic crystal is an artificial periodic dielectric medium
and exhibits a band structure in the photon dispersion like
the electronic band structure of 
solid~\cite{Yablonovitch1987a,John1987a,JJWM2nd}. 
The emission of photons is forbidden in the band gap because of
the null density of states. Hence, McQn is prohibited
provided that at least one of the photons involved in the process
falls in the band gap. 

In this paper, we study the rate of McQn, especially McQ3, in photonic 
waveguides in order to clarify the possibility of suppression
relative to RENP. We first describe the simplest photonic crystal
waveguide, namely a slab waveguide~\cite{YehYariv1976a} in 
Sec.~\ref{Sec:SlabWG}.
The band structure of a slab waveguide is shown and 
the Purcell factor~\cite{Purcell1946a}, which quantifies the suppression
or enhancement of emission in an environment, is introduced. 
In Sec.~\ref{Sec:BF}, we examine a photonic crystal waveguide of concentric 
stratified structure with a hollow core 
(Bragg fiber~\cite{YehYarivMarom1978a} in short) as a structure suitable
for RENP experiment. The rate of McQ3 in Bragg fiber is evaluated and
the specification required for sufficient suppression of McQ3 is shown
in Sec.~\ref{Sec:McQ3BF}.  Section \ref{Sec:SO} is devoted to 
summary and outlook.

\section{\label{Sec:SlabWG} Slab waveguide} 

\begin{figure}
 \centering
 \includegraphics[width=0.4\textwidth]{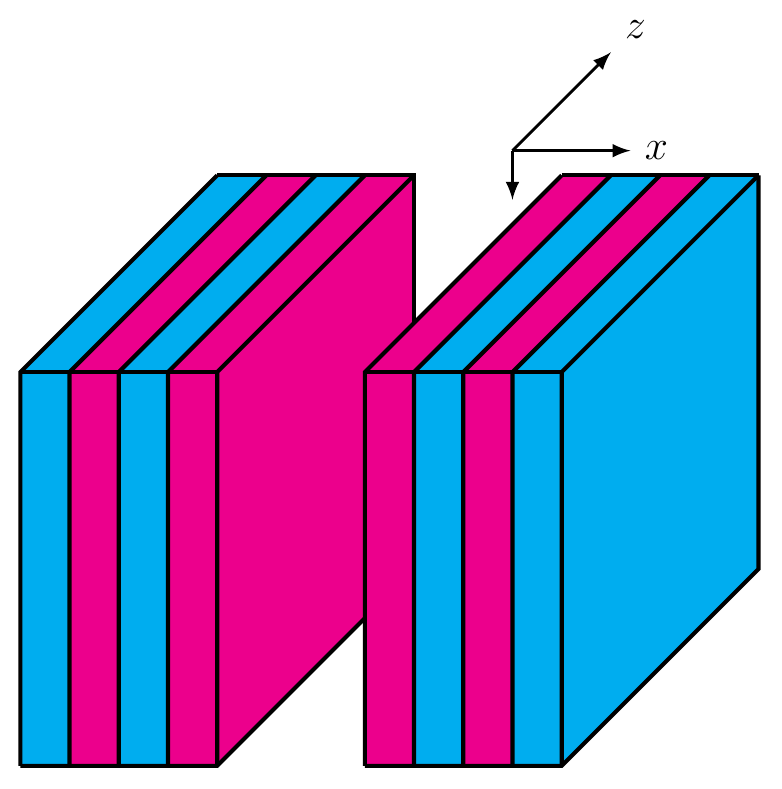}
 \caption{Slab waveguide.}
 \label{Fig:SlabWG3D}
\end{figure}

The suppression of photon emission is illustrated most 
simply by taking a nearly periodic system in one space dimension.
We thus describe the principle of manipulation of photon emissions in
photonic crystals taking a slab waveguide as an example.
A slab waveguide~\cite{YehYariv1976a} is a system of two confronted 
infinite-area slabs of stratified dielectric media as illustrated
in Fig.~\ref{Fig:SlabWG3D}.
The electromagnetic field is supposed to be confined in the
space between two slabs (the core) by Bragg reflection and
propagate to the $z$ direction in the core.

\subsection{\label{SubSec:SlabTF} Transfer matrix of a slab}

\begin{figure}
 \centering
 \includegraphics[width=0.5\textwidth]{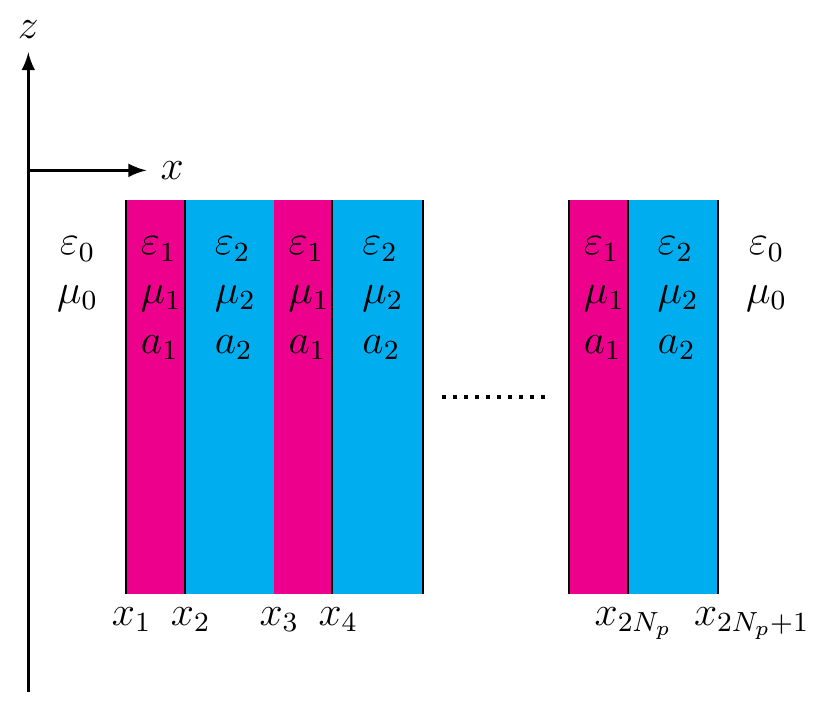}
 \caption{Specification of the slab.}
 \label{Fig:Slab}
\end{figure}

Each slab of the slab waveguide has a periodic structure of 
alternating refractive index
$n_i=\sqrt{\varepsilon_i\mu_i}$
($i=0,1,2$) as shown in Fig.~\ref{Fig:Slab}, where $i=0$ represents 
the regions of no dielectric medium,
and $i=1(2)$ corresponds to
the layer of thickness $a_{1(2)}$ made of the first (second) kind
of medium. We note that the permittivity $\varepsilon_i$
and the permeability $\mu_i$ are the relative ones, so that 
$\varepsilon_0=\mu_0=1$, and they are assumed real throughout this work.
The positions of the slab surfaces and the layer interfaces are denoted
by $x_j$ ($j=1,2,\cdots 2N_p+1$) with $N_p$ the number of layer pairs 
as indicated in Fig.~\ref{Fig:Slab}.

A propagating field of angular frequency $\omega$ and wave number 
$\beta$ in the $z$ direction is written as 
\begin{equation}
\psi(t,x,y,z)=\psi(x)e^{i(\beta z-\omega t)}\,,
\end{equation}
where $\psi(x)$ satisfies the one-dimensional homogeneous Helmholtz
equation 
\begin{equation}
(\partial_x^2+k_i^2)\psi(x)=0\,,\  
k_i=\sqrt{\varepsilon_i\mu_i\omega^2-\beta^2}\,.
\end{equation}
We assume the translational symmetry in the $y$ direction,
hence $y$ dependence is omitted.
As in ordinary metal waveguides, the fields in the slab waveguide are
classified into the transverse magnetic and electric (TM and TE)
modes~\cite{Jackson3rd}.  
For the TM (TE) mode, $\psi=E_z\ (H_z)$ and other components of fields 
are derived from $E_z(H_z)$, where $\bm{E}$ ($\bm{H}$) is 
the electric (magnetic) field.

The fields in each region or layer are given as follows:
\begin{equation}\label{Eq:CoreField}
 E_z(x)=A_0 e^{ik_0 x}+B_0 e^{-ik_0 x}\,,\
 H_z(x)=C_0 e^{ik_0 x}+D_0 e^{-ik_0 x}\,,
\end{equation}
in the region of $0<x<x_1$;
\begin{equation}\label{Eq:OddField}
 E_z(x)=A_{2j-1} e^{ik_1\xi}+B_{2j-1} e^{-ik_1\xi}\,,\
 H_z(x)=C_{2j-1} e^{ik_1\xi}+D_{2j-1} e^{-ik_1\xi}\,,
\end{equation}
for $x_{2j-1}<x<x_{2j}$ ($1\leq j\leq N_p$) with the local coordinate
$\xi$ defined by $x=x_{2j-1}+\xi$ 
($0<\xi<a_1$);
\begin{equation}\label{Eq:EvenField}
 E_z(x)=A_{2j} e^{ik_2\xi}+B_{2j} e^{-ik_2\xi}\,,\
 H_z(x)=C_{2j} e^{ik_2\xi}+D_{2j} e^{-ik_2\xi}\,,
\end{equation}
for $x_{2j}<x<x_{2j+1}$ ($1\leq j\leq N_p$) with $x=x_{2j}+\xi$ 
($0<\xi<a_2$);
\begin{equation}\label{Eq:OutField}
 E_z(x)=A_{2N_p+1} e^{ik_0\xi}+B_{2N_p+1} e^{-ik_0\xi}\,,\
 H_z(x)=C_{2N_p+1} e^{ik_0\xi}+D_{2N_p+1} e^{-ik_0\xi}\,,
\end{equation}
for $x_{2N_p+1}<x$ with $x=x_{2N_p+1}+\xi$.
We note that the coefficients $A$'s ($B$'s) and $C$'s ($D$'s) represent
the waves propagating toward the positive (negative) $x$ direction.

The fields in adjacent regions satisfy the continuity conditions of 
$E_z(x)$, $H_z(x)$ and their derivatives at the boundary. 
Such conditions are conveniently described by using transfer matrices
that relates the coefficients $A$'s, $B$'s, $C$'s and $D$'s in different
regions~\cite{YehYarivHong1977a}.
For the TM mode, we find the following connection formula,
\begin{equation}\label{Eq:CF}
 \begin{pmatrix}
  A_{2N_p+1} \\
  B_{2N_p+1}
 \end{pmatrix}
 =T_\text{TM}
  \begin{pmatrix}
   A_0 \\
   B_0
  \end{pmatrix}\,,
\end{equation}
where the transfer matrix is defined by
\begin{equation}\label{Eq:TM}
 T_\text{TM}:=T^{(e)}_\text{TM}T^{(1)}_\text{TM}
             (T^{(2)}_\text{TM}T^{(1)}_\text{TM})^{N_p-1}
             T^{(0)}_\text{TM}\,,
\end{equation}
with
\begin{equation}
 T^{(0)}_\text{TM}=
 \frac{1}{2}\begin{pmatrix}
             (1+\ell^{(0)}_\text{TM}) e^{i k_0 x_1} &
             (1-\ell^{(0)}_\text{TM}) e^{-i k_0 x_1}\\
             (1-\ell^{(0)}_\text{TM}) e^{i k_0 x_1} &
             (1+\ell^{(0)}_\text{TM}) e^{-i k_0 x_1}
            \end{pmatrix}\,,
 \quad \ell^{(0)}_\text{TM}=\frac{k_1\varepsilon_0}{\varepsilon_1 k_0}\,,
\end{equation}
\begin{equation}\label{Eq:TMi}
 T^{(i)}_\text{TM}=
 \frac{1}{2}\begin{pmatrix}
             (1+\ell^{(i)}_\text{TM}) e^{i k_i a_i} &
             (1-\ell^{(i)}_\text{TM}) e^{-i k_i a_i}\\
             (1-\ell^{(i)}_\text{TM}) e^{i k_i a_i} &
             (1+\ell^{(i)}_\text{TM}) e^{-i k_i a_i}
            \end{pmatrix}\,,
 \quad i=1,2\,,
 \quad\ell^{(1)}_\text{TM}=\frac{k_2\varepsilon_1}{\varepsilon_2 k_1}
      =\frac{1}{\ell^{(2)}_\text{TM}}\,,
\end{equation}
and
\begin{equation}
 T^{(e)}_\text{TM}=
 \frac{1}{2}\begin{pmatrix}
             (1+\ell^{(e)}_\text{TM}) e^{i k_2 a_2} &
             (1-\ell^{(e)}_\text{TM}) e^{-i k_2 a_2}\\
             (1-\ell^{(e)}_\text{TM}) e^{i k_2 a_2} &
             (1+\ell^{(e)}_\text{TM}) e^{-i k_2 a_2}
            \end{pmatrix}\,,
 \quad \ell^{(e)}_\text{TM}=\frac{k_0\varepsilon_2}{\varepsilon_0 k_2}\,.
\end{equation}
As for the TE modes, we obtain
\begin{equation}
 \begin{pmatrix}
  C_{2N_p+1} \\
  D_{2N_p+1}
 \end{pmatrix}
 =T_\text{TE}
  \begin{pmatrix}
   C_0 \\
   D_0
  \end{pmatrix}\,,
\end{equation}
where $T_\text{TE}$ is given by exchanging $\varepsilon_i$ and
$\mu_i$ in $T_\text{TM}$.
We note that $T_\text{TM}$ and $T_\text{TE}$ are unimodular;
$\det T_\text{TM}=\det T_\text{TE}$=1. 
An elementary derivation of transfer matrix is given in
Appendix \ref{App:DTM}.

\subsection{\label{SubSec:SlabBS} Band structure of the slab}
In the limit of large number of layer pairs, $N_p\to\infty$,
the surface effect can be neglected and the system has a strict periodicity
of period $a:=a_1+a_2$. Then, the field obeys $\psi(x+a)=e^{iKa}\psi(x)$
as is well known as Bloch's theorem in quantum mechanics. 
In terms of the field coefficients (of odd layers), 
Bloch's theorem is expressed as
\begin{equation}\label{Eq:Bloch}
 \begin{pmatrix}
  A_{2j+1} \\
  B_{2j+1}
 \end{pmatrix}
 =e^{iK_\text{TM}a}
  \begin{pmatrix}
   A_{2j-1} \\
   B_{2j-1}
  \end{pmatrix}\,,\ 
 \begin{pmatrix}
  C_{2j+1} \\
  D_{2j+1}
 \end{pmatrix}
 =e^{iK_\text{TE}a}
  \begin{pmatrix}
   C_{2j-1} \\
   D_{2j-1}
  \end{pmatrix}\,.
\end{equation}
While, as described above, $\psi(x)$ and $\psi(x+a)$ are related by the 
transfer matrix of unit period,
\begin{equation}\label{Eq:UnitTransfer}
 \begin{pmatrix}
  A_{2j+1} \\
  B_{2j+1}
 \end{pmatrix}
 =U_\text{TM}
  \begin{pmatrix}
   A_{2j-1} \\
   B_{2j-1}
  \end{pmatrix}\,,\ 
 \begin{pmatrix}
  C_{2j+1} \\
  D_{2j+1}
 \end{pmatrix}
 =U_\text{TE}
  \begin{pmatrix}
   C_{2j-1} \\
   D_{2j-1}
  \end{pmatrix}\,,
\end{equation}
where $U_X:=T^{(2)}_X T^{(1)}_X$ ($X=\text{TM, TE}$) is unimodular.

It is obvious from Eqs.~\eqref{Eq:Bloch} and \eqref{Eq:UnitTransfer}
that $e^{iK_X a}$ is an eigenvalue of $U_X$, that is
$e^{iK_X a}=h^X\pm\sqrt{(h^X)^2-1}$, where 
\begin{equation}
h^X:=\text{tr}\,U_X/2=\cos k_1 a_1\cos k_2 a_2-
                      \frac{\ell^{(1)}_X+\ell^{(2)}_X}{2}
                      \sin k_1 a_1\sin k_2 a_2\,.
\end{equation}
We see that $h^X$ is real if $k_1$ and $k_2$ are real, which is 
the case above the light line, that is $\omega\geq|\beta|$, 
provided that $n_{1,2}\geq 1$. 
Then, the nature of the Bloch wave is determined by the magnitude of $h^X$:
If $(h^X)^2<1$, $e^{iK_X a}=h^X\pm i\sqrt{1-(h^X)^2}$ is a complex
number of unit absolute value and thus $K_X$ is real, 
namely the field is oscillating in the $x$ direction. 
If $(h^X)^2>1$, $e^{iK_X a}$ is real and $K_X$ is pure imaginary
(modulo $\pi/a$), namely the field is exponentially dumping (or growing)
in the $x$ direction.  
In other words, the region of $(h^X)^2<1$ in the $\beta$-$\omega$ plane is 
the allowed band, and that of $(h^X)^2>1$ is the forbidden band, 
forming the band gap. 
Therefore, the band edge is determined by the condition
$(h^X)^2=1$~\cite{YehYarivHong1977a}.

\begin{figure}
 \centering
 \includegraphics[width=0.6\textwidth]{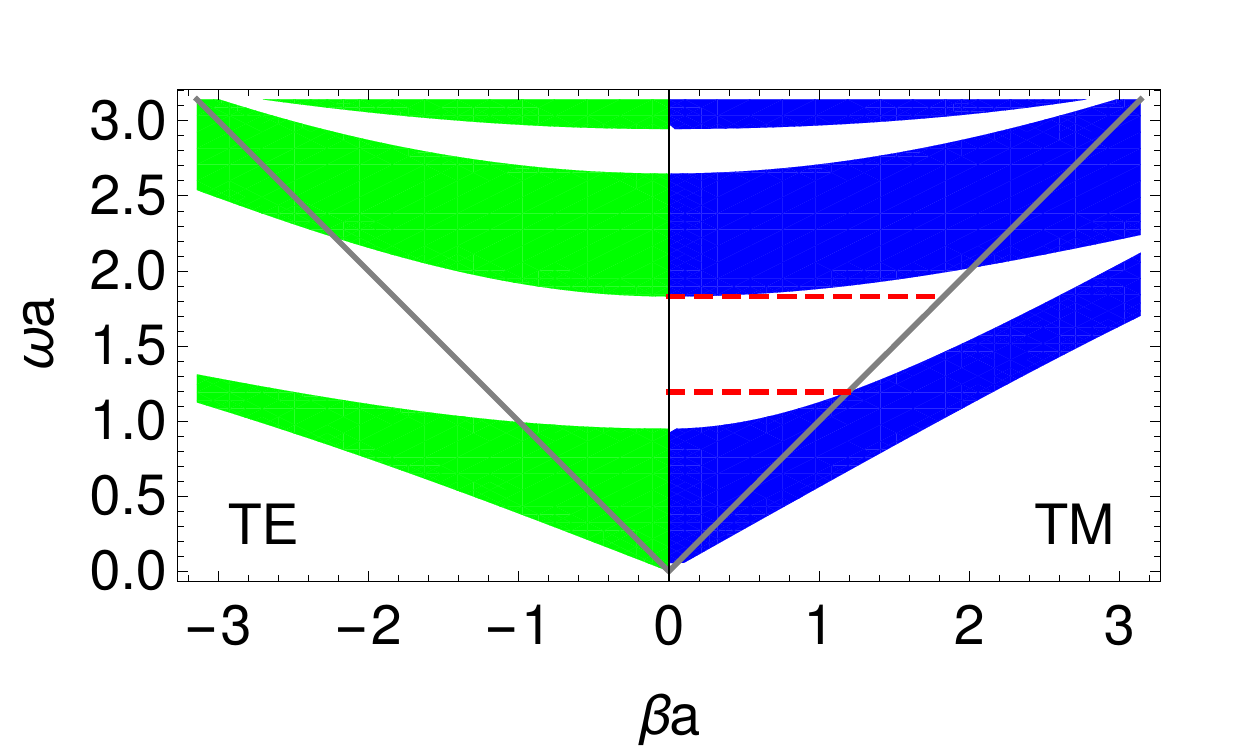}
 \caption{Band structure of a dielectric slab. The parameters of the slab
          are $n_1=4.6$, $n_2=1.6$, and 
          $\sqrt{n_1^2-1}\,a_1=\sqrt{n_2^2-1}\,a_2$.
          The blue (green) region is the allowed 
          band of the TM (TE) mode.}
 \label{Fig:Band}
\end{figure}

In Fig.~\ref{Fig:Band}, we illustrate the band structure of
a slab taking the example of dielectric media 
of $n_1=4.6$ (tellurium based glass)
and $n_2=1.6$ (polystyrene)~\cite{Fink1999a}. 
The thicknesses of layers are chosen to satisfy 
$\sqrt{n_1^2-1}\,a_1=\sqrt{n_2^2-1}\,a_2$,
which corresponds to the quarter-wave condition along the light line.
The permeabilities are assumed $\mu_{1,2}=1$.
We present the allowed band of the TM (TE) mode for
the positive (negative) $\beta$ as the blue (green) filled region. 
The diagonal solid lines show the light lines.
We observe that the region of $\omega$ between two red dashed lines 
is a complete band gap, namely no field exists in the slab for any 
$|\beta|(<\omega)$ in this region of $\omega$. 
Thus, in the complete band gap, the Bragg reflection in the slab
results in the perfect reflection for any incident angle at the surface of 
the (semi-infinite) slab~\cite{Winn1998a}, and a pair of slabs placed
face-to-face as shown in Fig.~\ref{Fig:SlabWG3D} would function as 
an optical waveguide.

It is useful in the following discussion of McQ3 suppression
to understand how the first band gap is determined.
For a quarter-wave stack, which satisfies $k_1 a_1=k_2 a_2=:\phi$, 
both the upper and lower boundaries of the first band gap are given by 
$h_X=-1$, namely
\begin{equation}\label{Eq:Boundaries}
\cos\phi=\pm\sqrt{\frac{(\ell^{(1)}_X+\ell^{(2)}_X)/2-1}
                       {(\ell^{(1)}_X+\ell^{(2)}_X)/2+1}}\,.
\end{equation}
We note that $\phi$ and $\ell^{(1,2)}_X$ are functions of $\beta$ and
$\omega$, and thus Eq.~\eqref{Eq:Boundaries} describes the boundary curves
of the band gap in the $\beta$-$\omega$ plane.

\subsection{\label{SubSec:SlabPurcell} Emission in the slab waveguide}
It is well known that the emission rate of photon in a cavity is modified 
from that in the free space because of the change in the density of 
photon states~\cite{Purcell1946a,Kleppner1981a,Goy1983a,Hulet1985a}. 
Similarly, photonic crystals may be used to suppress
optical emissions~\cite{Bykov1972a,Bykov1975a,Yablonovitch1987a,John1987a}. 
(For recent reviews, see e.g.~\cite{NodaFujitaAsano2007a,Pelton2015a}.) 
The degree of suppression of an emission process is quantified
by the ratio of the emission rate in an environment (a photonic crystal
waveguide in the present work) $\Gamma$ and that in the free space 
$\Gamma_\text{FS}$, $F_P:=\Gamma/\Gamma_\text{FS}$, which is called
as Purcell factor in the literature. 

\begin{figure}
 \centering
 \includegraphics[width=0.3\textwidth]{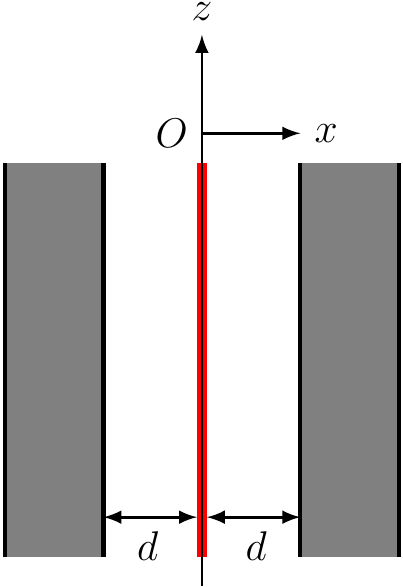}
 \caption{Slab waveguide with a source (in solid red).}
 \label{Fig:SWG2D}
\end{figure}

We consider the emission process from a localized source in the core of 
a slab waveguide as depicted in Fig.~\ref{Fig:SWG2D}. 
The core size is $2d$, the source uniform in the $y$-$z$ plane is placed
at the center of the core.
The slabs are of finite number of layer pairs ($N_p<\infty$)
as supposed in any realistic experiment, 
so that the prohibition in the band gap is not perfect, 
namely $F_P\neq 0$. 
Since the complete band gap is determined by the band gap of the TM mode
as seen in Fig.~\ref{Fig:Band}, we evaluate the Purcell factor of
the TM mode, $\psi(x)=E_z(x)$, in the following.

The field in the core in the presence of the source of unit strength
is described by the inhomogeneous Helmholtz equation,
\begin{equation}\label{Eq:IHHE}
(\partial_x^2+k_0^2)E_z(x)=-\delta(x)\,,
\end{equation}
where the origin of the $x$ coordinate is chosen to be the position of
the source.
One finds that the field is continuous at the position of the source
and its derivative has a discontinuity there:
\begin{equation}\label{Eq:ConDiscon}
E_z(+0)-E_z(-0)=0\,,\ 
\partial_x E_z(+0)-\partial_x E_z(-0)=-1\,.
\end{equation}

The field in the region of $0<x<d$ is expressed as
\begin{equation}\label{Eq:FPx}
E_z(x)=A_0 e^{i k_0 x}+B_0 e^{-i k_0 x}\,.
\end{equation}
For $-d<x<0$, the symmetry of the system under the flip of 
the $x$ coordinate dictates  
\begin{equation}\label{Eq:FNx}
E_z(x)=A_0 e^{-i k_0 x}+B_0 e^{i k_0 x}\,,
\end{equation}
so that the first equation in Eq.~\eqref{Eq:ConDiscon} is
automatically satisfied.
Substituting Eqs.~\eqref{Eq:FPx} and \eqref{Eq:FNx} into
the second equation of Eq.~\eqref{Eq:ConDiscon}, we obtain
\begin{equation}\label{Eq:CD}
A_0-B_0=\frac{i}{2 k_0}\,.
\end{equation}

Since the emitter at $x=0$ is only the source, no incoming
wave exists in the outside region of the slab waveguide.
This defines the outgoing wave condition (known as Sommerfeld 
radiation condition). In the region of $x>x_{2N_p+1}$ in 
Fig.~\ref{Fig:Slab}, this condition implies $B_{2N_p+1}=0$. 
Then the connection formula in Eq.~\eqref{Eq:CF} leads to
\begin{equation}\label{Eq:OGWC}
 \begin{pmatrix}
  A_{2N_p+1} \\
  0
 \end{pmatrix}
 =T_\text{TM}
  \begin{pmatrix}
   A_0 \\
   B_0
  \end{pmatrix}\,.
\end{equation}

The total outgoing flux is identified as
\begin{equation}
P=\frac{\omega\varepsilon_0}{k_0}|A_{2N_p+1}|^2\,.
\end{equation}
It is straightforward to solve $A_{2N_p+1}$ from Eqs.~\eqref{Eq:CD} and
\eqref{Eq:OGWC},
\begin{equation}\label{Eq:FluxSWG}
A_{2N_p+1}=\frac{i}{2 k_0}
           \frac{1}{(T_\text{TM})_{21}+(T_\text{TM})_{22}}\,,
\end{equation}
where we have used the unimodularity of $T_\text{TM}$.
The flux in the free space $P_\text{FS}$ is given by putting 
$T_\text{TM}=I$, where $I$ denotes the identity matrix.
We thus find the expression of the Purcell factor in terms of
the transfer matrix $T_\text{TM}$ in Eq.~\eqref{Eq:TM},
\begin{equation}\label{Eq:PFSWG}
F_P=\frac{P}{P_\text{FS}}
   =\frac{1}{|(T_\text{TM})_{21}+(T_\text{TM})_{22}|^2}\,.
\end{equation}

\begin{figure}
 \centering
 \includegraphics[width=0.7\textwidth]{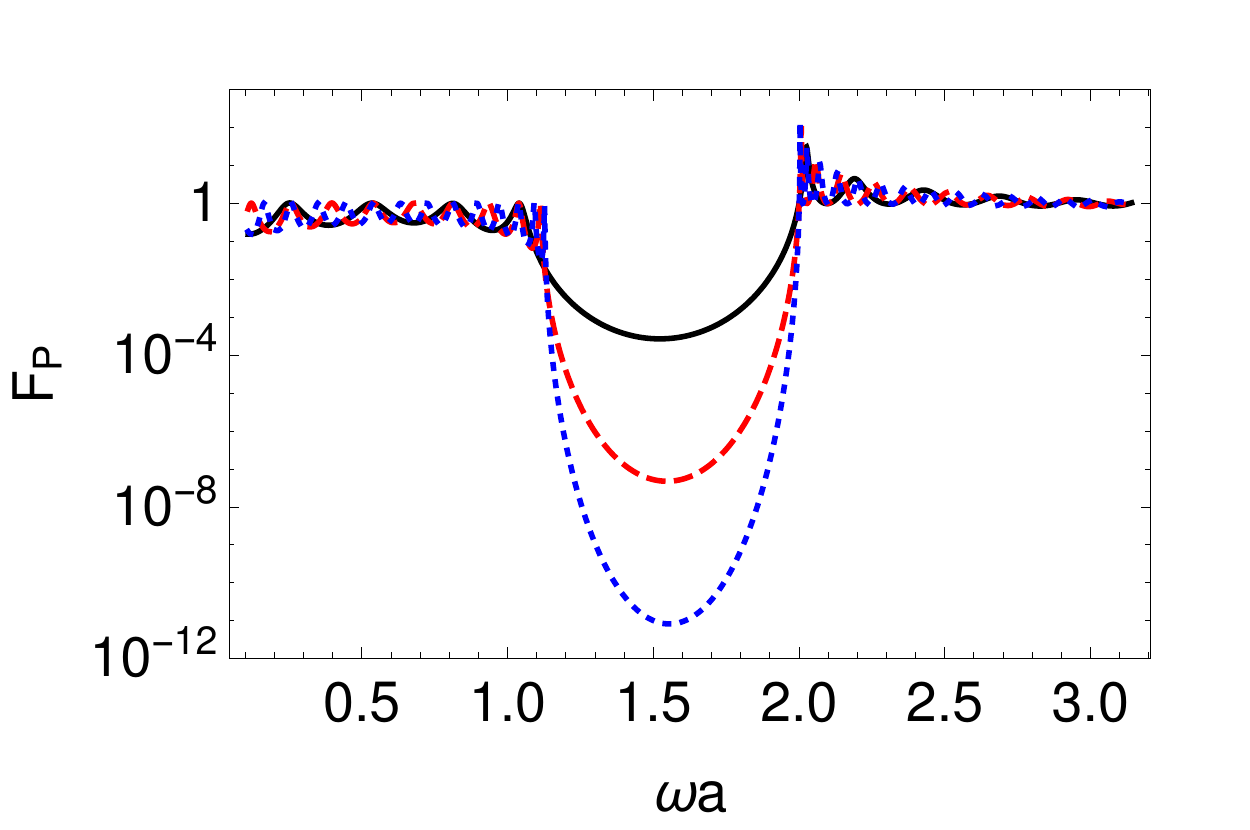}
 \caption{Purcell factor near the light line, $\beta a=\omega a-0.1$, 
          in the slab waveguide of core size $2d=2.5a$. Other parameters
          are the same as Fig.~\ref{Fig:Band}. The solid black, dashed red
          and dotted blue lines represent $N_p=5$, 10 and 15 respectively.}
 \label{Fig:PFSWG}
\end{figure}

In Fig.~\ref{Fig:PFSWG}, we illustrate the Purcell factor of the slab 
waveguide whose parameters are specified by the core size $2d=2.5a$, 
the refractive indices $n_{1,2}$ and the thicknesses of layers
$a_{1,2}$ being the same as Fig.~\ref{Fig:Band}, and the number of
layer pairs $N_p=5$ (solid black), 10 (dashed red) and 15 (dotted blue).
Given these parameters, the Purcell factor is a function of the frequency
$\omega$ and the wave number $\beta$. We choose the line 
$\beta a=\omega a-0.1$, which is near the light line, motivated by
the kinematics of McQ3 discussed in Sec.~\ref{Sec:McQ3BF}.
The first band gap is clearly seen.
Furthermore, we observe that the Purcell factor in the band gap decreases
exponentially as $N_p$ increases. This behavior is analytically
confirmed for the large $N_p$ limit in Appendix~\ref{App:LargeN}.

\section{\label{Sec:BF} Bragg fiber}
\begin{figure}
 \centering
 \raisebox{5ex}{\includegraphics[width=0.3\textwidth]{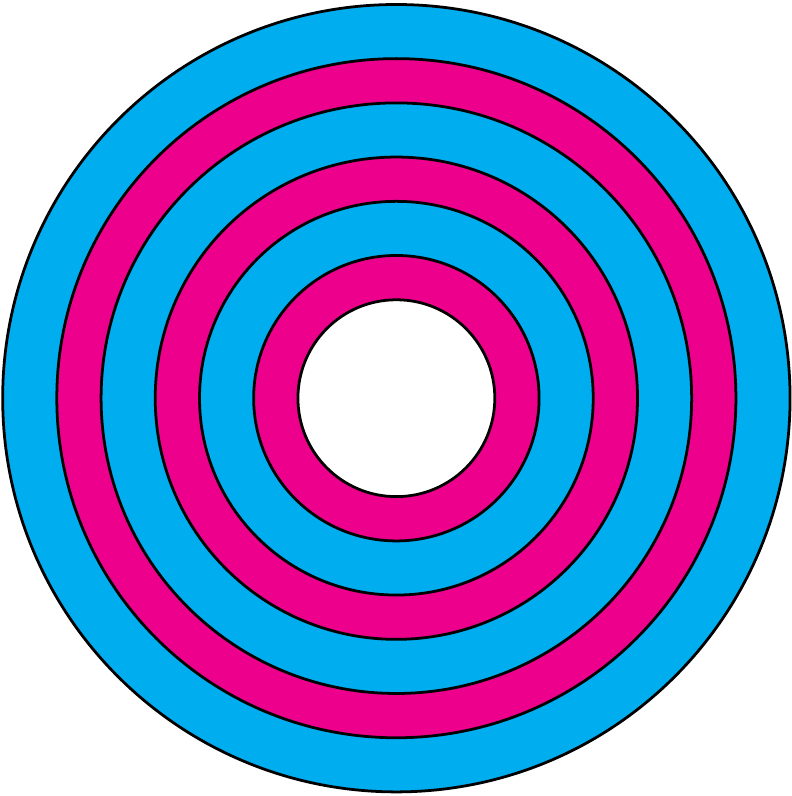}}
 \hspace{0.1\textwidth} 
 \includegraphics[width=0.4\textwidth]{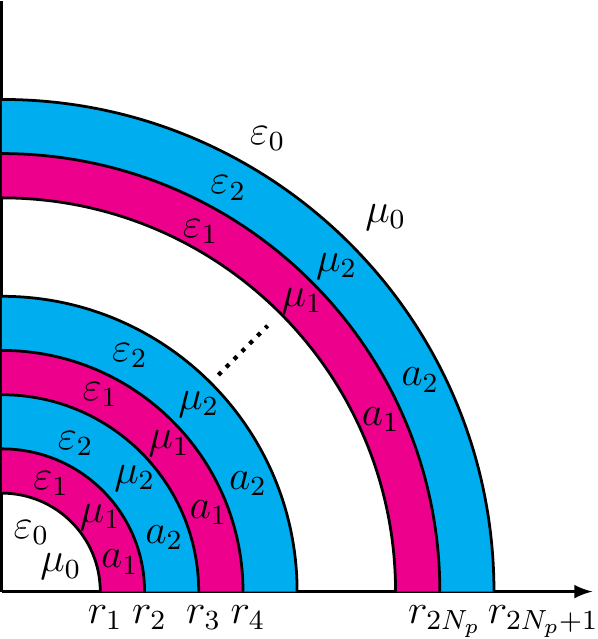}
 \caption{Bragg fiber. Left: The whole cross section. Right: 
          Specification of the Bragg fiber of alternating 
          structure of layer pairs of thickness $a_i$ ($i=1,2$)
          with the relative electric (magnetic) constant 
          $\varepsilon_i$ ($\mu_i$).
          The $z$ direction is taken to be the fiber axis.}
 \label{Fig:BF}
\end{figure}

Although the slab waveguide discussed above captures the essential idea
of emission rate suppression in a photonic crystal waveguide, it is not 
realistic enough because of its unclosed core (in the $y$ direction). 
In this section, instead, we consider the Bragg 
fiber~\cite{YehYarivMarom1978a,Johnson2001a,Ibanescu2003a}, 
which has a closed cross section as is depicted in Fig.~\ref{Fig:BF}.

The fields propagating in the $z$ direction are described in 
the cylindrical coordinate as
\begin{equation}
\psi(t,r,\theta,z)=\psi(r,\theta)e^{i(\beta z-\omega t)}\,,
\end{equation}
where $\psi(r,\theta)$ satisfies the two-dimensional Helmholtz
equation, $(\bm{\nabla}_t^2+k_i^2)\psi(r,\theta)=0$.
We take $\psi=E_z$ and $H_z$ independent so that the transverse 
($r$ and $\theta$) components are given in terms of 
the derivatives of the $z$ components.

\subsection{\label{SubSec:BFTF} Transfer matrix of the Bragg fiber}
The fields in the region of $r_j<r<r_{j+1}$ 
are expressed as
\begin{align}
\label{Eq:EzHankel}
E_z(r,\theta)&=\left[A_j H^{(1)}_m(k_j r)+B_j H^{(2)}_m(k_j r)\right]
               e^{i m\theta}\,,\\
\label{Eq:HzHankel}
H_z(r,\theta)&=\left[C_j H^{(1)}_m(k_j r)+D_j H^{(2)}_m(k_j r)\right]
               e^{i m\theta}\,,
\end{align}
where $H^{(1,2)}_m$ are Hankel functions; $A_j$, $B_j$, $C_j$ and $D_j$ 
are complex coefficients; and $k_j=k_{1(2)}$ for odd (even) $j$
in the range $1\leq j\leq 2N_p$.
The region of $j=0$, that is $0=r_0<r<r_1$, is the hollow core,
where the transverse wave number is given by $k_0$;
the region of $r_{2N_p+1}<r$ ($j=2N_p+1$) represents the exterior of
the fiber, where $k_{2N_p+1}=k_0$.
We note that the Hankel functions $H^{(1)}_m$ and $H^{(2)}_m$ 
represent the outgoing and incoming wave respectively.

The field coefficients $\bm{v}_j:=(A_j,B_j,C_j,D_j)^\intercal$ in 
the adjacent regions are related by the connection formula at
the interface,
\begin{equation}
M_j(r_j)\bm{v}_j=M_{j-1}(r_j)\bm{v}_{j-1}\,,
\end{equation}
where 
\begin{equation}\label{Eq:CM}
M_j(r)=\begin{pmatrix}
        H^{(1)}_m(k_j r) & H^{(2)}_m(k_j r) & 0 & 0\\
        -\frac{\omega\varepsilon_j}{\beta k_j}H^{(1)\prime}_m(k_j r) &
        -\frac{\omega\varepsilon_j}{\beta k_j}H^{(2)\prime}_m(k_j r) &
        \frac{m}{k_j^2 r}H^{(1)}_m(k_j r) &             
        \frac{m}{k_j^2 r}H^{(2)}_m(k_j r) \\             
        0 & 0 & H^{(1)}_m(k_j r) & H^{(2)}_m(k_j r) \\
        \frac{m}{k_j^2 r}H^{(1)}_m(k_j r) &             
        \frac{m}{k_j^2 r}H^{(2)}_m(k_j r) &
        -\frac{\omega\mu_j}{\beta k_j}H^{(1)\prime}_m(k_j r) &
        -\frac{\omega\mu_j}{\beta k_j}H^{(2)\prime}_m(k_j r) \\
       \end{pmatrix}\,,
\end{equation}
$\varepsilon_j=\varepsilon_{1(2)}$ for odd (even) $j$ in the range
$1\leq j\leq 2N_p$, $\varepsilon_j=\varepsilon_0$ for $j=0,2N_p+1$,
and the same for $\mu_j$.
The primed Hankel functions denote the derivatives with respect to
their own argument.
Thus, the exterior coefficients are expressed in terms of the core 
coefficients using the transfer matrix $T$,
\begin{equation}\label{Eq:BFCF}
\bm{v}_{2N_p+1}=T\bm{v}_0\,,
\end{equation}
where
\begin{equation}
T:=T_{2N_p+1}T_N\cdots T_2T_1\,,
\end{equation}
and
\begin{equation}\label{Eq:TBF}
T_j:=\left[M_j(r_j)\right]^{-1}M_{j-1}(r_j)\,.
\end{equation}
The explicit formula of $T_j$ is given in Appendix~\ref{App:TBF}.
We note that $E_z$ and $H_z$ have to be finite at $r=0$,
which implies $A_0=B_0$ and $C_0=D_0$.

\subsection{\label{SubSec:BFBS} Band structure of the Bragg fiber}
The cladding of the Bragg fiber is not strictly periodic even 
in the limit of $N_p\to\infty$. 
It, however, can be regarded as periodic in the asymptotic region, 
$k_jr\gg 1$ since the curvature is negligible \cite{XuLeeYariv2000a}.
Thus the Bragg fiber is expected to exhibit a band structure for
sufficiently large $N_p$.

One finds that $M_j(r)$ in Eq.~\eqref{Eq:CM} becomes block diagonal and 
the TM and TE modes are separated in the asymptotic region as in the
slab case. Moreover, it turns out, using the asymptotic forms of
the Hankel functions, 
\begin{equation}
H^{(1,2)}_m(kr)\simeq\sqrt{\frac{2}{\pi kr}}\,e^{\pm i[kr-(2m+1)\pi/4]}\,,
\end{equation}
that the field coefficients follow Eq.~\eqref{Eq:UnitTransfer} with
$U_X$ of the same diagonal elements as the slab regardless the value
of $m$ provided that an appropriate phase redefinition of 
the coefficients is made in order to absorb the difference between
the local coordinate employed in the formulation of the slab and 
the global one in the Bragg fiber. 
Since the band structure is solely determined by the trace of $U_X$ 
as discussed in Sec.~\ref{SubSec:SlabBS}, the Bragg fiber has the identical
band structure to the slab.

\subsection{\label{SubSec:BFPurcell} Purcell factor of the Bragg fiber}
The emission from a source $s(r,\theta)$ (uniform in the $z$ direction)
in the Bragg fiber is described by the inhomogeneous two-dimensional 
Helmholtz equation,
$(\bm{\nabla}_t^2+k_i^2)\psi(r,\theta)=s(r,\theta)$, where one may write
$\psi(r,\theta)=\psi(r)e^{im\theta}$ and $s(r,\theta)=s(r)e^{im\theta}$.
We consider a localized source of unit strength at the center of the core 
($r=0$), namely $s(r,\theta)=-\delta(r)/(2\pi r)$. This implies $m=0$,
and the transfer matrix becomes block diagonal,
\begin{equation}
T=\begin{pmatrix}
   T_\text{TM} & 0 \\
   0 & T_\text{TE}
  \end{pmatrix}\,,
\end{equation}
as seen in Eq.~\eqref{Eq:CM}. The TM and TE modes are separated as in
the case of the slab waveguide.

The solution of the inhomogeneous Helmholtz equation is given by 
the Green's function with the boundary condition of outgoing wave, 
$G(r)=(i/4)H^{(1)}_0(k_0 r)$.
Applying this solution to the TM mode, from which the complete band gap
is deduced, we obtain the core field,
\begin{equation}
E_z(r)=\left(A_0+\frac{i}{4}\right)H^{(1)}_0(k_0 r)+A_0 H^{(2)}_0(k_0 r)\,,
\end{equation}
as well as the exterior field,
\begin{equation}
E_z(r)=A_{2N_p+1} H^{(1)}_0(k_0 r)\,,
\end{equation}
where $A_0=B_0$ and the outgoing wave condition $B_{2N_p+1}=0$ are used.

The flux at the exterior surface ($r>r_{2N_p+1}$) is given by
\begin{equation}
 P=\frac{2\omega\varepsilon_0}{k_0^2}|A_{2N_p+1}|^2\,.
\end{equation}
Solving the TM part of Eq.~\eqref{Eq:BFCF},
\begin{equation}
\begin{pmatrix}
A_{2N_p+1} \\
0
\end{pmatrix}
=T_\text{TM}\begin{pmatrix}
             A_0+\frac{i}{4}\\
             A_0
            \end{pmatrix}\,,
\end{equation}
with the help of the unimodularity of $T_\text{TM}$, we obtain
\begin{equation}
A_{2N_p+1}=\frac{i}{4}\,\frac{1}{(T_\text{TM})_{21}+(T_\text{TM})_{22}}\,.
\end{equation}
In the free space, $T_\text{TM}$ is the unit matrix, hence $A_{2N_p+1}=i/4$.
Finally, we obtain the expression of the Purcell factor of the Bragg fiber as
\begin{equation}\label{Eq:PFBF}
F_P=\frac{P}{P_\text{FS}}
   =\frac{1}{|(T_\text{TM})_{21}+(T_\text{TM})_{22}|^2}\,.
\end{equation}
We note that this expression has the same form as the case of the slab
waveguide in Eq.~\eqref{Eq:PFSWG} but the content of $T_\text{TM}$ of
the Bragg fiber is different from that of the slab in general.

\begin{figure}
 \centering
 \includegraphics[width=0.7\textwidth]{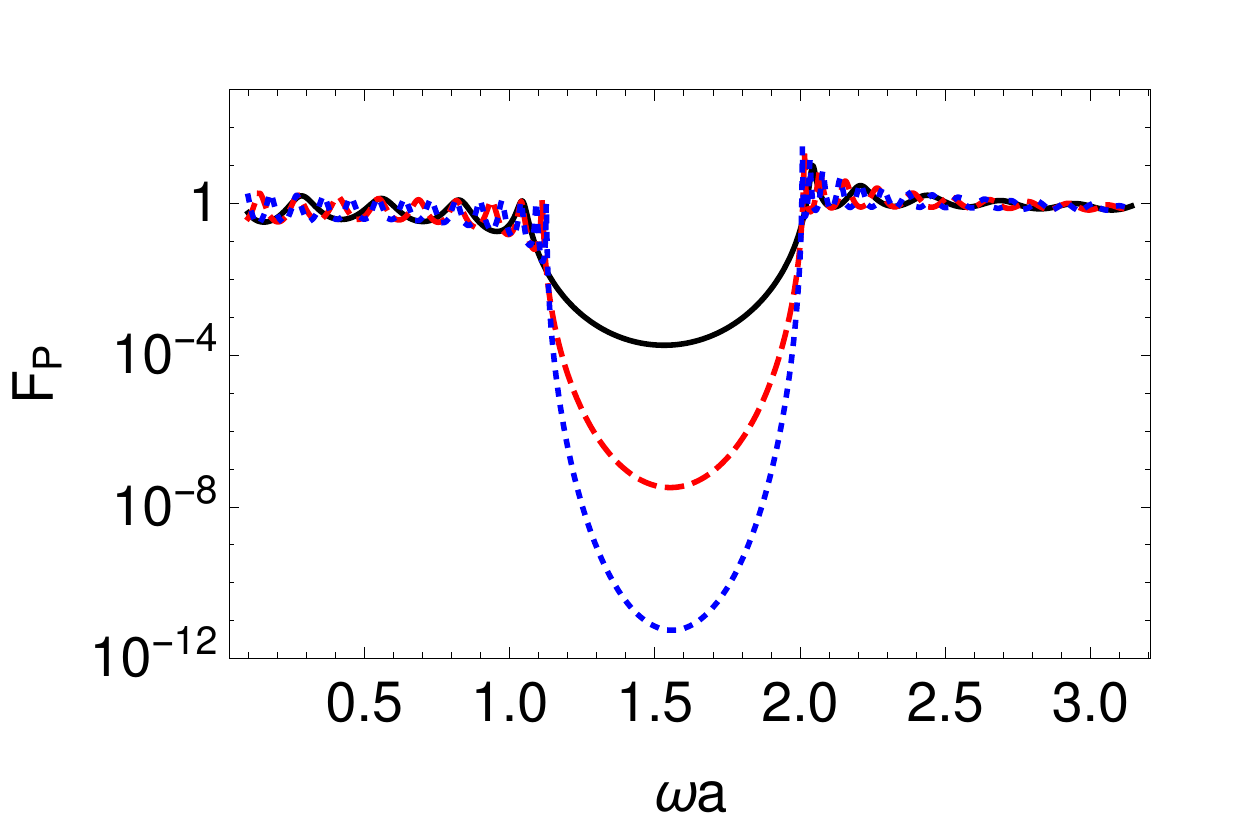}
 \caption{Purcell factor of the Bragg fiber near the light line,
          $\beta a=\omega a-0.1$. The core radius is $r_1=2a$
          and other parameters are the same as Figs.~\ref{Fig:Band}
          and \ref{Fig:PFSWG}.}
 \label{Fig:PFBF}
\end{figure}

We present an example of the numerical evaluation of the Purcell factor
in the Bragg fiber in Fig.~\ref{Fig:PFBF}.
The refractive indices $n_{1,2}$ and the thicknesses of layers $a_{1,2}$
are the same as those in Fig.~\ref{Fig:Band}, and the core radius is 
taken as $r_1=2a$.
The solid black, dashed red and dotted blue lines represent $N_p=5$, 
10 and 15 respectively.
The position and the depth of the band gap is practically the same as
the slab waveguide of the same layer structure and the Purcell factor
in the band gaps exponentially decreases as the number of the layer pairs
increases, similarly to the slab waveguide. Hence, the McQn process
in which at least one of the photons is emitted in the band gap can be
strongly suppressed in the Bragg fiber.

\section{\label{Sec:McQ3BF} Suppression of McQ3 in the Bragg fiber}
McQ3 has the largest rate among McQn's if it is allowed by the parity.
In this section, we evaluate its rate in the Bragg fiber. 
We study a few combinations of refractive indices in order to clarify
the necessary condition for the sufficient background suppression compared
to the RENP signal.

An experiment using the Xe gas target filled in the core of an Bragg fiber
is taken as an illustration. 
The Xe gas is prepared in a macrocoherent
state \cite{Fukumi2012a} and
the trigger laser irradiated along the fiber axis for the RENP process 
also induces the McQ3 background. As is mentioned in Sec.~\ref{Sec:Intro}
the deexcitation from $|e\rangle=6\text{s}\,^3\text{P}_1$ of 8.437 eV 
excitation energy to the the ground state $|g\rangle=5\text{p}\,^1\text{S}_0$
is considered.

\subsection{McQ3 rate in the free space}
The free-space rate of the above Xe McQ3 process is explained in detail 
in Ref.~\cite{YoshimuraSasaoTanaka2015a}.
To be self-contained, here we summarize the relevant formulae.

The differential spectral rate is given by 
\begin{equation}
\frac{d\Gamma_\text{FS}}{d\omega_1}=
\frac{\Gamma_0}{\omega_0}|D|^2\omega_1^2\omega_2^2\,,
\end{equation}
where $\Gamma_0$ is an overall rate, which is irrelevant here,
$\omega_0$ denotes the angular frequencies of the trigger light,
$\omega_{1,2}$ are those of emitted photons, the energy denominator
$D$ is written as
\begin{align}
D(\omega_1)=&\frac{1}{\omega_{pe}+\omega_0}
              \left(\frac{1}{\omega_{eg}-\omega_1}+
                    \frac{1}{\omega_0+\omega_1}\right)+
             \frac{1}{\omega_{pe}+\omega_1}
              \left(\frac{1}{\omega_{eg}-\omega_0}+
                    \frac{1}{\omega_0+\omega_1}\right)\nonumber\\
            &+\frac{1}{\omega_{pg}-\omega_0-\omega_1}
               \left(\frac{1}{\omega_{eg}-\omega_0}+
                     \frac{1}{\omega_{eg}-\omega_1}\right)\,,
\end{align}
and $\hbar\omega_{eg}=8.437\ \text{eV}$ is the excitation energy.
We note that $\omega_1$ and $\omega_2$ are not independent once
the trigger frequency is specified owing to the energy conservation,
$\omega_{eg}=\omega_0+\omega_1+\omega_2$.
The spectral rate is obtained by integrating over $\omega_1$,
\begin{equation}
\Gamma_\text{FS}=
\int_{\omega_{eg}/2-\omega_0}^{\omega_{eg}/2}d\omega_1
 \frac{d\Gamma_\text{FS}}{d\omega_1}\,.
\end{equation}

The magnitude of the longitudinal (with respect to the trigger) momentum
of each emitted photon, which is necessary to evaluate the Purcell factor
in the following, is dictated by the energy-momentum conservation
satisfied in atomic processes with the macrocoherence;
\begin{equation}
\beta_i=\omega_i|\cos\theta_i|\,,\ i=1,2\,,
\end{equation}
where
\begin{equation}
\cos\theta_i=\frac{\omega_{eg}}{\omega_0}-1-
             \frac{\omega_{eg}(\omega_{eg}-2\omega_0)}{2\omega_0\omega_i}\,.
\end{equation}

\subsection{McQ3 rate in the Bragg fiber}
The rate of McQ3 is modified in the Bragg fiber and the modification is
described by the Purcell factor discussed above. The differential rate
in the Bragg fiber is represented as
\begin{equation}\label{Eq:dGam}
 \frac{d\Gamma_\text{BF}}{d\omega_1}=
 \frac{d\Gamma_\text{FS}}{d\omega_1}
 F_p(\omega_0,\beta_0) F_p(\omega_1,\beta_1) F_p(\omega_2,\beta_2)\,,
\end{equation}
where each Purcell factor is explicitly shown as a function of
the corresponding photon's $\omega$ and $\beta$.
The Purcell factor for the trigger is omitted below
because it is common for RENP (signal) and McQ3 (background).%
\footnote{%
It is evident that the density of neutrino states is hardly 
affected by the Bragg fiber (or photonic crystals in general). 
The trigger photon in RENP (and McQ3), however, is subject to the Purcell
factor $F_p(\omega_0,\beta_0)$. Although the $S/N$ ratio is independent
of it, the absolute rate of RENP does depend on $F_p(\omega_0,\beta_0)$.
It is required to choose the structure of the photonic crystal waveguide
and the trigger parameters $(\omega_0,\beta_0)$ such that 
$F_p(\omega_0,\beta_0)$ is not strongly suppressed in the frequency range
relevant to RENP.}
Thus, the degree of the background suppression by the Bragg fiber 
is given by
\begin{equation}\label{Eq:rBFFS}
r_\text{BF/FS}=
 \frac{1}{\Gamma_\text{FS}} 
 \int_{\omega_{eg}/2-\omega_0}^{\omega_{eg}/2}d\omega_1
  \frac{d\Gamma_\text{FS}}{d\omega_1}
  F_p(\omega_1,\beta_1)F_p(\omega_2,\beta_2)\,.
\end{equation}

\begin{figure}
 \centering
 \includegraphics[width=0.7\textwidth]{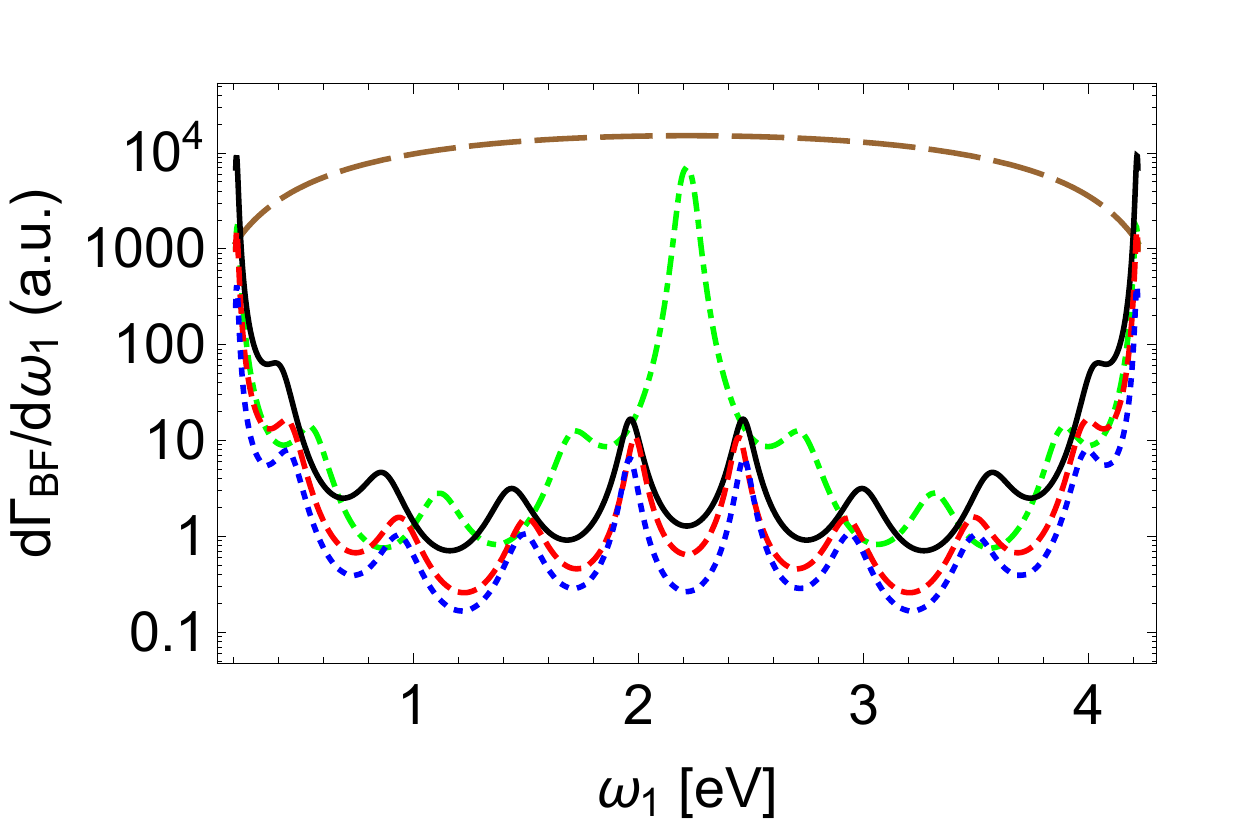}
 \caption{The differential rate of McQ3 in the Bragg fiber of $N_p=5$
          (in an arbitrary unit).
          The trigger frequency is $\omega_0=0.95 \omega_{eg}/2$.
          The dot-dashed green, solid black, dashed red and dotted blue
          lines show the cases of $(n_1,n_2)=(4.6,1.6)$, 
          (4.6,1.2), (4.8,1.3) and (5.0,1.3) respectively.
          The free-space spectrum is also shown in the long-dashed brown line
          for comparison.} 
 \label{Fig:dGam}
\end{figure}

In Fig.~\ref{Fig:dGam}, we present the shape of McQ3 differential rate
in the Bragg fiber given in Eq.~\eqref{Eq:dGam} for several combinations
of refractive indices $n_1$ and $n_2$. 
The structure of the Bragg fiber follows the quarter-wave stack condition
along the light line, namely $\sqrt{n_1^2-1}\,a_1=\sqrt{n_2^2-1}\,a_2$, 
as in Sec.~\ref{SubSec:BFPurcell};
while the period $a(=a_1+a_2)$ is optimized to minimize $r_\text{BF/FS}$.
We employ the number of layer pairs of $N_p=5$ as an illustration.
The trigger frequency is chosen as $\omega_0=0.95 \omega_{eg}/2$ 
so that $\omega_1+\omega_2$ is slightly larger than $\omega_{eg}/2$
as used in Ref.~\cite{YoshimuraSasaoTanaka2015a}. 
We note that the transverse length scale $a$ of the Bragg fiber is
of order $1/\omega_{eg}\simeq 0.15\ \mu\text{m}$, which is much smaller
than the scale of the metal waveguide in Sec.~\ref{Sec:Intro},
and the supposed neutrino mass range below is 
$\sim \omega_{eg}/4\simeq 2\ \text{eV}$ or less.

We observe that the rate suppression is significant for relatively larger
index differences, 
$(n_1,n_2)=(4.6,1.2)$, (4.8,1.3) and (5.0,1.3).
In these cases, the band gap is wide enough such that at least one of
the emitted photons is in the band gap owing to the energy conservation.
In the case of the combination $(4.6,1.6)$ employed in the fabrication
of Bragg fiber in Ref.~\cite{Fink1999a}, however, the band gap is not
sufficiently wide and there is an energy range in which both of the photons
are in the allowed band as seen as the peak around $\omega_1\simeq 2.2$ eV.

In the case of infinite periodic structure described in 
Sec.~\ref{SubSec:SlabBS}, the boundaries of the band gap are determined
by Eq.~\eqref{Eq:Boundaries}. 
Since the kinematics of McQ3 under present discussion requires 
$\omega\sim\beta$, the band gap near the light line is relevant.
The crossing points of the light line and the boundaries of 
the first band gap are given by
\begin{equation}
\omega_\pm a=
 \frac{\sqrt{n_1^2-1}+\sqrt{n_2^2-1}}{\sqrt{n_1^2-1}\sqrt{n_2^2-1}}\,
 \arccos\left(\pm\sqrt{\frac{(\ell^{(1)}_\text{TM}+\ell^{(2)}_\text{TM})/2-1}
                            {(\ell^{(1)}_\text{TM}+\ell^{(2)}_\text{TM})/2+1}}
        \right)\,,
\end{equation}
and
\begin{equation}
\frac{\ell^{(1)}_\text{TM}+\ell^{(2)}_\text{TM}}{2}
=\frac{1}{2}\left(\frac{n_1^2\sqrt{n_2^2-1}}{n_2^2\sqrt{n_1^2-1}}+
                  \frac{n_2^2\sqrt{n_1^2-1}}{n_1^2\sqrt{n_2^2-1}}\right)\,.
\end{equation}
We note that $\omega_{+(-)}$ gives the lower (upper) boundary of the
first band gap along the light line. We find that at least one of the
emitted photons has an energy in the band gap and the McQ3 process is 
prohibited if $2\omega_+<\omega_-$, provided that the period $a$ is 
chosen to satisfy $\omega_1+\omega_2=\omega_-$. 
If the band gap is not wide enough such that this inequality
is not satisfied, both photons are emitted in the first allowed band
in a part of the phase space and McQ3 is not prohibited. 
As seen in Fig.~\ref{Fig:dGam}, this is the case for $(n_1,n_2)=(4.6,1.6)$.

\begin{figure}
 \centering
 \includegraphics[width=0.7\textwidth]{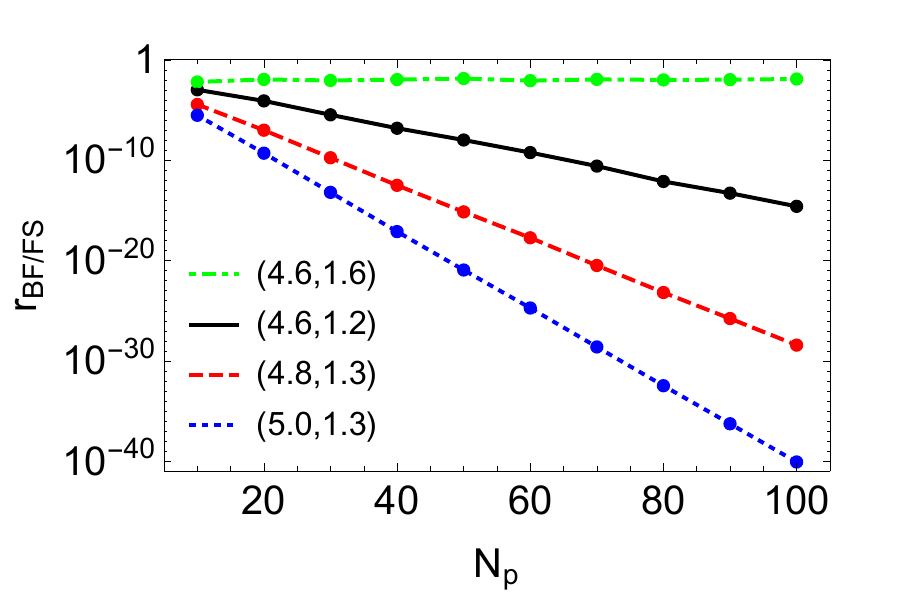}
 \caption{McQ3 suppression factor in the Bragg fiber.
          The pairs of refractive indices $(n_1,n_2)$ are chosen as
          indicated and the trigger frequency is 
          $\omega_0=0.95 \omega_{eg}/2$.}
 \label{Fig:rBFFS}
\end{figure}

Figure \ref{Fig:rBFFS} shows $r_\text{BF/FS}$ as a function of $N_p$. 
Other parameters that define the structure of Bragg fiber and
the trigger frequency are the same as Fig.~\ref{Fig:dGam}.
We find that, as is expected, the dependence of $r_\text{BF/FS}$ on
$N_p$ is approximately exponential for larger index differences.
It turned out, on the other hand, that the rate suppression is not
significant and practically independent of $N_p$ for the combination 
of (4.6,1.6). This is due to the existence of the energy range
in which both the photons are emitted in the allowed band as mentioned
above. 

Hence the required suppression of McQ3
is attainable with the Bragg fiber of finite $N_p$ provided that 
a wide band gap is realized with a sufficiently large index difference
and $N_p$ is large enough.

\section{\label{Sec:SO} Summary and outlook}
We have studied the macrocoherently enhanced multiphoton process
that is a potential source of serious background in the atomic neutrino
process RENP. The proposal of background suppression using photonic
crystals is examined in detail. The essential idea is the prohibition
of photon emission in the band gap of photonic crystals. 

Employing the method of transfer matrix, the periodic dielectric 
structure of slab layers that has alternating refractive indices is
shown to exhibit such a band structure because of the Bragg reflection
at the interfaces of two dielectric media. 
The suppression of emission in the slab waveguide made of two finite slabs
is quantified by the Purcell factor. We find that the Purcell factor
decreases exponentially as the number of layer pairs $N_p$ increases.

Then, as a more realistic experimental setup, the McQ3 process in
the Bragg fiber is discussed. The Bragg fiber possesses practically 
the identical band structure and Purcell factor to the slab waveguide.
The degree of McQ3 background suppression is given by $r_\text{BF/FS}$
in Eq.~\eqref{Eq:rBFFS}, which compares the McQ3 rate in the Bragg fiber
to that in the free space.
We have numerically evaluated $r_\text{BF/FS}$ using the transfer matrix
and found that it also decreases exponentially as $N_p$ increases for
sufficiently large index differences. 

The Bragg fiber fabricated in the laboratory is made of dielectric media 
of $n_1=4.6$ and $n_2=1.6$ in the infrared region~\cite{Fink1999a}. 
It was found that the suppression factor $r_\text{BF/FS}$ does not
scale exponentially for this pair of refractive indices because of 
the relatively narrower band gap. Our numerical results indicate
that a larger index contrast such as (4.6,1.2) 
is desirable for the background rejection. 
The polymer of index $\sim 1.3$ in the optical region is commercially 
available (e.g. CYTOP by AGC). 
To further reduce the index toward $\sim 1.2$,
one may replace part of the material with a gas like aerogels.

Even if the synthesis of material of required indices
in the optical region is impossible, there are two possible ways to overcome.
One is to find a target or an experimental setup which energy scale is
low enough. Since the refractive index tends to be larger for lower
frequencies, the necessary index difference could be obtained 
with increased $n_1$.
Another is to consider more elaborated structures than those discussed
in the present work. For example, a set of slabs of periods $a$ and $2a$
may be combined. The lowest band gap of the latter locates at the upper
half of the first allowed band of the former, so that their combination
effectively exhibits a wider band gap. Both of these improvement are
under investigation.

The higher-order macrocoherent QED processes McQn 
($n\geq 4$) are also potentially dangerous. 
In the case that McQ3 is allowed by the parity, the next
allowed is McQ5 of five photons. 
As the concrete evaluation of the McQ5 rate is not available at present, 
we roughly estimate the ratio of McQ5 and McQ3 rates by 
counting the coupling and phase-space factors as 
$\alpha^2/((4\pi)^4 2!3!)\sim O(10^{-10})$, 
where $\alpha$ is the fine structure constant. In addition to the
native rate suppression, the photon veto is expected more effective
for higher McQn. If a background photon emitted along the Bragg fiber
is detected with 99.9\% efficiency, the McQ5 rate is virtually 
reduced by $(10^{-3})^4=10^{-12}$. Combining two suppression factors,
McQ5 (or higher) seems rather harmless. (The photon veto is also helpful
to mitigate the requirement for McQ3 suppression.)

To design a realistic experiment of RENP,
the method to excite the atoms/molecules in a macroscopic target to 
an appropriate state for the neutrino emission should be studied
as well as the background processes. 
In the PSR experiments in Refs.~\cite{Miyamoto2014a,Miyamoto2015a},
the initial coherent state of macroscopic target is prepared by
the Raman process with two parallel lasers propagating in the same
direction. It turns out that a pair of counter-propagating lasers
is necessary for the emission of massive neutrinos unlike the
PSR case in which the emitted particles are massless photons.
A PSR experiment of para-hydrogen target with counter-propagating lasers
is in preparation at Okayama University in order to show that 
the macrocoherent emission of massive particles is possible.

In conclusion, the suppression mechanism of McQ3 background in RENP 
by the photonic crystal waveguide works in principle. Other backgrounds
of higher orders are relatively suppressed and the photon veto is 
effective for them. Further research and development works are
necessary to design a background-free RENP experiment.

\section*{Acknowledgments}
This work is supported in part by JSPS KAKENHI Grant Numbers 
JP25400257 (MT), 15H02093 (MT, NS and MY), 
15K13468 (NS) and 16H00868 (KT).

\appendix

\section{\label{App:DTM}
         Derivation of transfer matrix}
The boundary condition of the electromagnetic field at an interface of
two media is prescribed by Maxwell equations. The condition necessary to 
derive the transfer matrices is that the tangential components of
$\bm{E}$ and $\bm{H}$ are continuous at the interface.
Hence, in the case of the slab waveguide in 
Sec.~\ref{Sec:SlabWG}, the tangential components of the fields satisfy
the boundary conditions $E_{y,z}(x_i-0)=E_{y,z}(x_i+0)$ and 
$H_{y,z}(x_i-0)=H_{y,z}(x_i+0)$ at the interface of $x=x_i$ 
($1\leq i\leq 2N_p+1$).

It also follows from Maxwell equations that the transverse 
(not tangential) components of the fields are given by derivatives
of the longitudinal (i.e. $z$) components as is mentioned in 
Sec.~\ref{SubSec:SlabTF}. For the slab waveguide, one finds 
\begin{equation}
E_y(x)=-\frac{i}{k_i^2}\omega\mu_i\frac{\partial H_z}{\partial x}\,,\ 
H_y(x)=\frac{i}{k_i^2}\omega\varepsilon_i\frac{\partial E_z}{\partial x}\,,
\end{equation}
where $i=0,1$ or 2. Thus the derivatives of $E_z(x)$ and $H_z(x)$
as well as $E_z(x)$ and $H_z(x)$ themselves are continuous 
at the interface.

Applying the above result to the interface at $x=x_{2j}$ ($1\leq j\leq N_p$)
of the slab waveguide, one obtains
\begin{equation}
 A_{2j-1}e^{ik_1 a_1}+B_{2j-1}e^{-ik_1 a_1}=A_{2j}+B_{2j}\,,
\end{equation}
from the continuity of $E_z$, and
\begin{equation}
 \frac{\varepsilon_1}{k_1}(A_{2j-1}e^{ik_1 a_1}-B_{2j-1}e^{-ik_1 a_1})
 =\frac{\varepsilon_2}{k_2}(A_{2j}-B_{2j})\,,
\end{equation}
from the derivative of $E_z$ (i.e. $H_y$). A matrix representation
of these equations gives the transfer matrix $T^{(1)}_\text{TM}$ in
Eq.~\eqref{Eq:TMi}. The other transfer matrices in the main text are
obtained in the same manner.

\section{\label{App:LargeN}\boldmath%
         Large \texorpdfstring{$N_p$}{Np} behavior of the Purcell factor}
We use the Chebyshev identity,
\begin{equation}\label{Eq:CI}
U^n=\frac{1}{\lambda_+-\lambda_-}
    \left[(\lambda_+^n-\lambda_-^n)U-(\lambda_+^{n-1}-\lambda_-^{n-1})I
    \right]\,,
\end{equation}
where $U$ is a unimodular two-by-two matrix whose eigenvalues are 
$\lambda_\pm$.
Since $U_\text{TM}:=T^{(2)}_\text{TM}T^{(1)}_\text{TM}$ is unimodular,
the transfer matrix in Eq.~\eqref{Eq:TM} is represented as
\begin{equation}
 T_\text{TM}=
 (\lambda_+^{N_p-1}-\lambda_-^{N_p-1})S_1
 -(\lambda_+^{N_p-2}-\lambda_-^{N_p-2})S_0\,,
\end{equation}
where
\begin{equation}
S_j:=\frac{1}{\lambda_+-\lambda_-}
     T^{(e)}_\text{TM}T^{(1)}_\text{TM}U_\text{TM}^j T^{(0)}_\text{TM}
     \quad (j=0,1)\,,
\end{equation}
is independent of $N_p$ and the eigenvalues $\lambda_\pm$ of $U_\text{TM}$
are given by
\begin{equation}
\lambda_\pm=h^\text{TM}\pm\sqrt{(h^\text{TM})^2-1}\,,\ 
h^\text{TM}:=\frac{1}{2}\text{tr}\,U_\text{TM}\,.
\end{equation}

The elements of the transfer matrix $T_\text{TM}$ are given by
\begin{align}
(T_\text{TM})_{22}&=(\lambda_+^{N_p-1}-\lambda_-^{N_p-1})a_1
                    -(\lambda_+^{N_p-2}-\lambda_-^{N_p-2})a_0\,,\\
(T_\text{TM})_{21}&=(\lambda_+^{N_p-1}-\lambda_-^{N_p-1})b_1
                    -(\lambda_+^{N_p-2}-\lambda_-^{N_p-2})b_0\,,
\end{align}
where
\begin{equation}
S_j=\begin{pmatrix}
     a_j^* & b_j^*\\
     b_j   & a_j
    \end{pmatrix}\,.
\end{equation}
The inequality $(h^\text{TM})^2>1$ is satisfied in the band gap
as is described in Sec.~\ref{SubSec:SlabBS}.
For $h^\text{TM}>1$, we find $\lambda_+>1>\lambda_->0$ and
\begin{align}
(T_\text{TM})_{22}&\simeq\lambda_+^{N_p-2}(\lambda_+ a_1-a_0)\,,\\
(T_\text{TM})_{21}&\simeq\lambda_+^{N_p-2}(\lambda_+ b_1-b_0)\,,
\end{align}
as $N_p\to\infty$. For $h^\text{TM}<-1$, $\lambda_-<-1<\lambda_+<0$ and
\begin{align}
(T_\text{TM})_{22}&\simeq -\lambda_-^{N_p-2}(\lambda_- a_1-a_0)\,,\\
(T_\text{TM})_{21}&\simeq -\lambda_-^{N_p-2}(\lambda_- b_1-b_0)\,.
\end{align}
Thus we obtain 
\begin{equation}
F_P\simeq\frac{|\lambda_\pm|^{-2(N_p-2)}}
              {|\lambda_\pm(a_1+b_1)-(a_0+b_0)|^2}
\propto e^{-2N_p\log|\lambda_\pm|}\,,\ h^\text{TM}\gtrless\pm 1\,,
\end{equation}
for large $N_p$, i.e. $F_P$ exponentially decreases as $N_p$ increases.

\section{\label{App:TBF}\boldmath%
         Explicit expression of \texorpdfstring{$T_j$}{Tj}}
We present the matrix elements of $T_j$ defined in Eq.~\eqref{Eq:TBF}:
\begin{align}
(T_j)_{11}&=(T_j)_{22}^*\nonumber\\
&=i\frac{\pi}{4}k_j r_j
   \left[H^{(1)}_m(k_{j-1}r_j)H^{(2)'}_m(k_j r_j)
         -\frac{\varepsilon_{j-1}k_j}{\varepsilon_j k_{j-1}}
         H^{(1)'}_m(k_{j-1}r_j)H^{(2)}_m(k_j r_j)\right]\,,
\end{align}
\begin{align}
(T_j)_{12}&=(T_j)_{21}^*\nonumber\\
&=i\frac{\pi}{4}k_j r_j
   \left[H^{(2)}_m(k_{j-1}r_j)H^{(2)'}_m(k_j r_j)
         -\frac{\varepsilon_{j-1}k_j}{\varepsilon_j k_{j-1}}
         H^{(2)'}_m(k_{j-1}r_j)H^{(2)}_m(k_j r_j)\right]\,,
\end{align}
\begin{align}
(T_j)_{13}=(T_j)_{24}^*
=i\frac{\pi}{4}\frac{m\beta}{\omega\varepsilon_j}
  \left(\frac{k_j^2}{k_{j-1}^2}-1\right)
  H^{(1)}_m(k_{j-1}r_j)H^{(2)}_m(k_j r_j)\,,
\end{align}
\begin{align}
(T_j)_{14}=(T_j)_{23}^*
=i\frac{\pi}{4}\frac{m\beta}{\omega\varepsilon_j}
  \left(\frac{k_j^2}{k_{j-1}^2}-1\right)
  H^{(2)}_m(k_{j-1}r_j)H^{(2)}_m(k_j r_j)\,,
\end{align}
\begin{align}
(T_j)_{31}=(T_j)_{42}^*=(T_j)_{13}\big|_{\varepsilon\leftrightarrow\mu}\,,
\end{align}
\begin{align}
(T_j)_{32}=(T_j)_{41}^*=(T_j)_{14}\big|_{\varepsilon\leftrightarrow\mu}\,,
\end{align}
\begin{align}
(T_j)_{33}=(T_j)_{44}^*=(T_j)_{11}\big|_{\varepsilon\leftrightarrow\mu}\,,
\end{align}
\begin{align}
(T_j)_{34}=(T_j)_{43}^*=(T_j)_{12}\big|_{\varepsilon\leftrightarrow\mu}\,.
\end{align}

\bibliographystyle{utphys}
\bibliography{PSRRENP,PhotonicCrystals}

\end{document}